# PHYSICS of SUPERNOVAE: THEORY, OBSERVATIONS, UNRESOLVED PROBLEMS[*]

D. K. Nadyozhin

Institute for Theoretical and Experimental Physics, Moscow. nadezhin@itep.ru

The main observational properties and resulting classification of supernovae (SNe) are briefly reviewed. Then we discuss the progress in modeling of two basic types of SNe – the thermonuclear and core-collapse ones, with special emphasis being placed on difficulties relating to a consistent description of thermonuclear flame propagation and the detachment of supernova envelope from the collapsing core (a nascent neutron star). The properties of the neutrino flux expected from the core-collapse SNe, and the lessons of SN1987A, exploded in the Large Magellanic Cloud, are considered as well.

## 1. Basic Properties of Supernovae

During the past two decades, there has occurred a technological break-through in astronomy. The sensitivity, spectral and angular resolution of detectors in the whole electromagnetic diapason from gamma-rays through radio wavelengths has been considerably increased. Hence, it became possible to derive from the observed supernova spectra precise data on dynamics of supernova expansion and on composition and light curves of SNe as well. Moreover, neutrino astronomy overstepped the bounds of our Galaxy to discover the neutrino signal from SN1987A. This epoch-making event had a strong impact on the theory of supernovae.

   The main indicator for supernova *astronomical* classification is the amount of hydrogen observed in the SN envelopes. The Type I SNe (subtypes Ia, Ib, Ic) contain no hydrogen in their envelopes. The SN Ia are distinguished by strong Si lines in their spectra and form quite a homogeneous group of objects. On the contrary, Type II SNe (subtypes IIP, IIn, IIL, IIb) have clear lines of hydrogen in their spectra (Fig. 1). Figure 2 shows the light curves of several individual SNe – representatives of the SN types listed above. One can observe a wide variety of the light curves for Type II SNe. For instance, Type II SNe with linear light curves (subtype IIL) are brighter at their maximum light than SN 1987A (apparently of subtype IIP with a plateau-shaped light curve) by ~5 stellar magnitude – i.e. by a factor of 100!

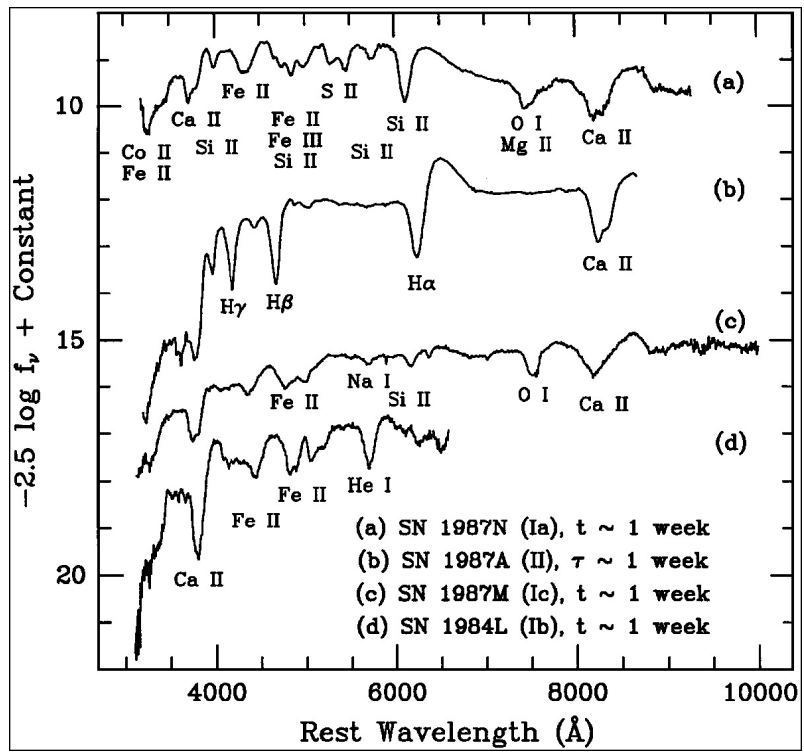

Figure 1. Spectra of SNe of different types at about one week after observed B-band maximum for SN 1987N (Ia), SN 1987M (Ic), SN 1984L (Ib) and after core collapse for SN 1987A (IIP) [1]. The spectral lines are broadened owing to the high velocities of the ejecta accelerated to several thousand km/s.

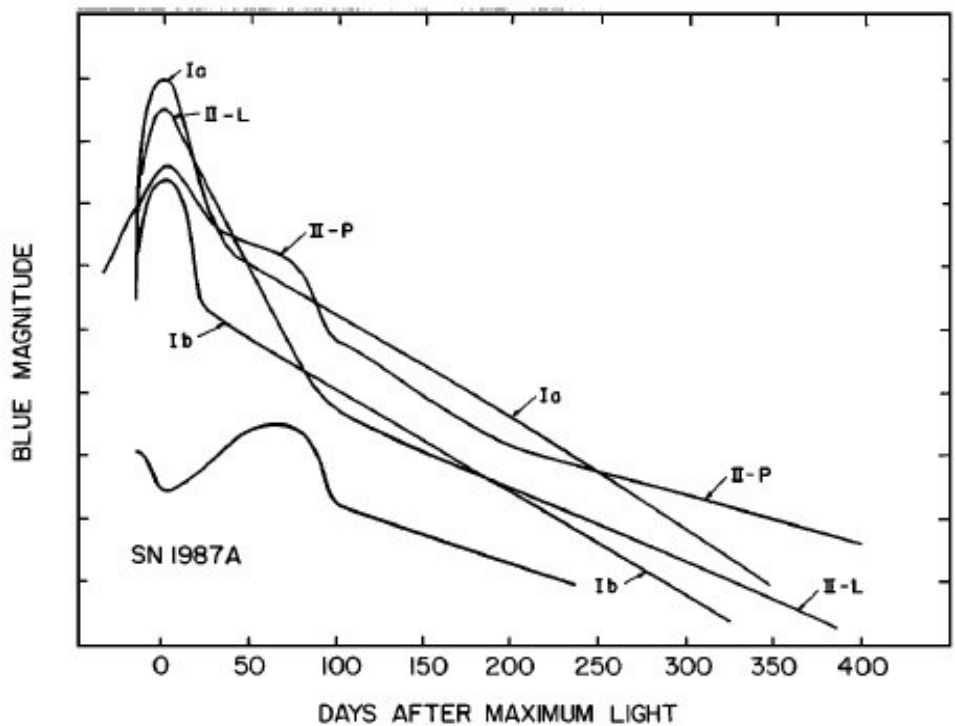

Figure 2. Schematic light curves for SNe of Types Ia, Ib, II-L, II-P, and SN 1987A [1,2]. The curve marked as SNe Ib is essentially an average of SNe Ib and SNe Ic.

*Physically*, there are two fundamental types of SNe: the thermonuclear SNe and the core-collapse ones, represented by Type Ia SNe (SN Ia) and by Type II, Ib, and Ic SNe, respectively (Fig. 3). The core-collapse SNe are subdivided into several subtypes depending on the amount of hydrogen hanging around the stellar core just before it begins to collapse. The progenitors of Type IIP and IIn SNe retain most of hydrogen envelope, typically as much as $(10 - 15)$ $M_\odot$. SNe IIn have in addition some hydrogen in an extended atmosphere formed by stellar wind on the top of dense hydrogen enriched envelope. The progenitors of Type IIL SNe have much less hydrogen, about $(0.1 - 1)$ $M_\odot$ whereas the spectra of Type IIb SNe show only traces of hydrogen during the first few days after the explosion, then these SNe become similar to SNe Ib. Types Ib and Ic progenitors virtually have no hydrogen left. Type Ic differs from Ib by lack of helium. The Type Ic progenitors are believed to have lost not only all hydrogen but a fair amount of their helium as well.

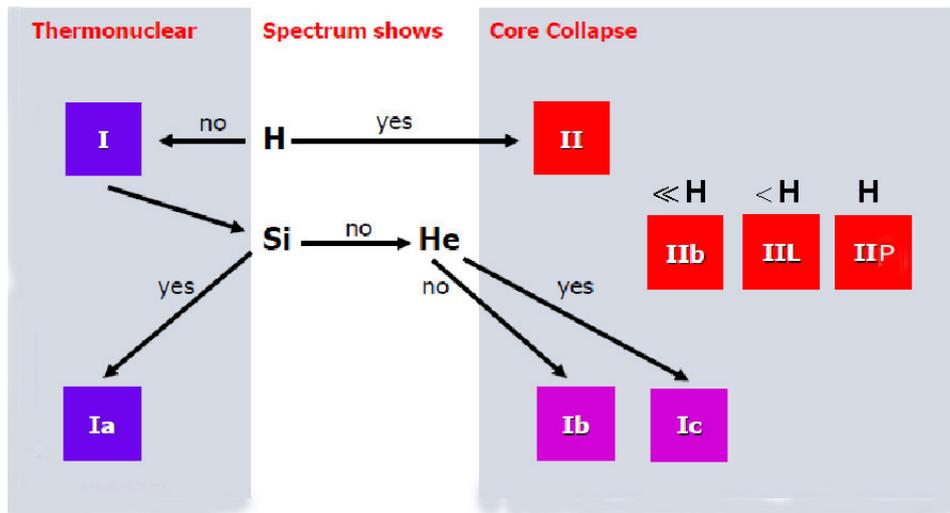

Figure 3. Empirical and physical classification scheme of supernovae. Adapted from [3].

**Thermonuclear SNe** (Type Ia) are believed to arise in a close binary system from explosive carbon burning either in a degenerate carbon-oxygen (CO) white dwarf as soon as due to accretion it's mass increases to a certain close to *Chandrasekhar mass* ($M \approx 1.4$ $M_\odot$) value or in the process of mergence of two white dwarf binary components. The explosion energy $E_{exp} \approx 10^{51}$erg comes from the thermonuclear burning of $^{12}$C and $^{16}$O mixture into $^{56}$Ni that has the maximum binding energy among nuclei with equal numbers of neutrons and protons. The white dwarf turns out to be totally disrupted in the explosion — no stellar remnant is left! The total energy of electromagnetic radiation $E_{rad} \approx 6 \times 10^{49}$erg is mostly supplied by the $^{56}$Ni $\rightarrow$ $^{56}$Co $\rightarrow$ $^{56}$Fe beta-decay. The hydrodynamic modeling of the SN Ia light curves shows that only a small fraction of the explosion energy is transformed into radiation: $E_{rad} / E_{exp} \approx 0.06$. This fraction is a factor of 2 less than the ratio of the energy released per $^{56}$Ni $\rightarrow$ $^{56}$Fe beta decay (5.5 MeV) to the energy produced per synthesized $^{56}$Ni (47 MeV for a CO mixture of equal C and O mass fraction).This happens because a portion of beta decay energy is expended on the hydrodynamic expansion of supernova debris. Hence, almost all $E_{exp}$ resides in the kinetic energy of envelope expanding with the mean velocity of $\approx 8\,000$ km/s. A comparison of the SN Ia models with observations shows that about $(0.6 - 1)$ M$_\odot$ of $^{56}$Ni is produced per SN Ia outburst.

**Core-collapse SN** outbursts are triggered by the gravitational collapse of the "iron" core of a mass $M_{Fe} = (1.3-2)$ M$_\odot$ into a neutron star. The gravitational binding energy of the nascent neutron star is about $(10-15)$ % of the core rest mass $M_{Fe}c^2$ and radiated in the form of neutrinos and antineutrinos of all the flavors (e,$\mu$,$\tau$). The total energy carried away amounts to $E_\nu = (3 - 5) \times 10^{53}$ erg depending on the core mass $M_{Fe}$. The explosion energy (kinetic energy of the envelope expansion) turns out to be three orders of magnitude less than $E_\nu$, $E_{exp} = (0.5-2) \times 10^{51}$ erg. It comes from the shock wave (SW) that is created at the boundary between a new-born neutron star and the envelope to be expelled. The total energy of electromagnetic radiation is as small as that for SN Ia, $E_{rad} = (1-10) \times 10^{49}$ erg, being strongly sensitive to the radius of the progenitor and the mass of its hydrogen envelope. This is the reason why the SN II light curves show a wide variety of shape and magnitude (Fig. 2). The light curve is powered by SW thermal energy with subsequent recombination of hydrogen and helium. The innermost layers of core-collapse SN envelope prove to be enriched with some amount of radioactive nuclei such as $^{56}$Ni [typically varying within $(0.02 - 0.2)$ M$_\odot$ for individual SNe] and also with less abundant $^{44}$Ti and $^{57}$Co. The heat generated by decay of these nuclides prevents a rapid decrease of SN luminosity owing to an adiabatic cooling. In particular, this explains long "tails" of the SN IIP light curves at time t > 100 days (Fig. 2). Only massive stars with M > $(8 - 10)$ M$_\odot$ can form iron cores and finish their life as core-collapse SNe. In late 1960's the hydrodynamic theory of SNe had drawn a conclusion that just before the explosion the most luminous hydrogen rich SNe II should have a red supergiant structure with radius of $(300 - 1000)$ R$_\odot$ whereas the blue supergiants of radius of $(30 - 100)$ R$_\odot$ could be appropriate for the explanation of much dimmer SNe. This prediction was remarkably confirmed by the observation of SN 1987A in the Large Magellanic Cloud and by recent inspection of prediscovery field of nearby SNe [4−6].

## 2. Thermonuclear flame

The ignition of thermonuclear fuel and propagation of the flame in degenerate stellar matter is a fundamental problem still to be solved to understand the basic mechanism of SN Ia explosions and finally to calibrate SN Ia as standard candles for observational cosmology. From the beginning the flame propagates by means of a sub-sonic deflagration being governed by the electron thermal conduction. The burning front proves to be extremely thin and fraught with a number of instabilities such as Rayleigh–Taylor, Kelvin–Helmholtz and Landau–Darries ones. Since the Reynolds number typically reaches value of the order of $10^{14}$, the front gets a strongly wrinkled structure and the burning becomes of turbulent nature. Owing to the growth of surface area covered by the front the rate of combustion considerably increases. As a result, after a time the deflagration can develop into a super-sonic detonation that is driven by a shock wave which heats matter up to the ignition. The transition from the deflagration to detonation is required in order to achieve the compliance between theoretically predicted chemical compositions of the SN Ia ejecta and that observed in the SN Ia spectra. However, for a group of discovered recently subluminous SNe Ia the deflagration alone seems to be adequate [7]. There is also a problem with understanding how and where the nuclear fuel actually begins to burn. The flame may flare up not necessarily in the very centre but either in a bubble somewhere off center or in separate little spots randomly distributed around. Figure 4 shows an example of a three-dimensional simulation of deflagration triggered by several dozens of burning spots chosen as initial condition. An extensive study of turbulent burning in degenerate matter of white dwarfs is under way. The current results and further references can be found in [7–15].

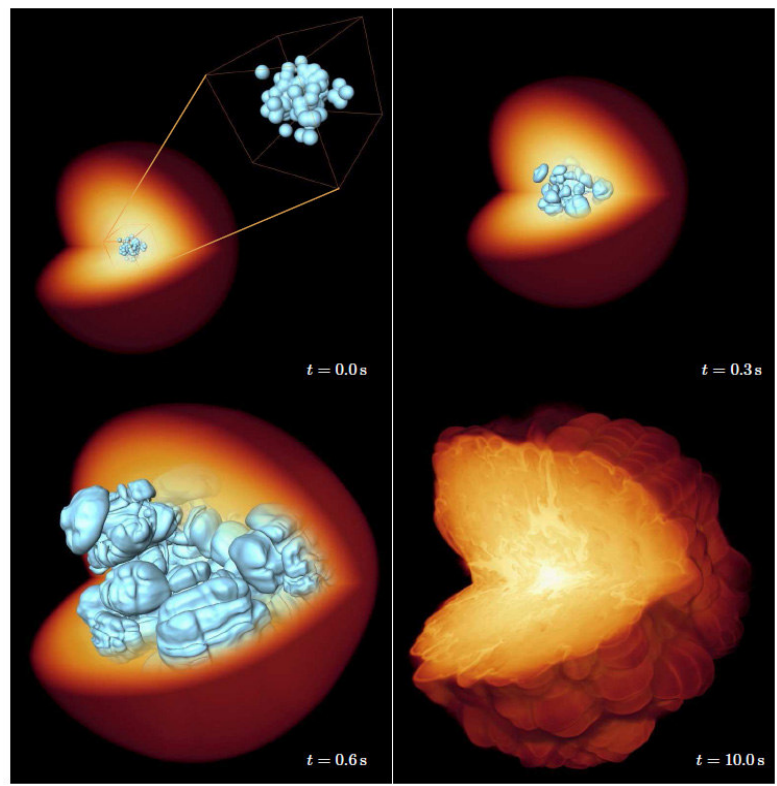

Figure 4. Multi-spot ignition scenario of the thermonuclear flame [8]. The level surfaces of density are shown in logarithmic scale. The last panel corresponds to the end of the calculation and is not on scale (actual dimension of the white dwarf star is considerably larger at t=10s). At this moment the burning is absolutely completed and the stellar material is in a state of free inertial expansion. In total about 0.4 $M_\odot$ of $^{56}$Ni was produced. The explosion energy was about $8\times10^{50}$ erg.

## 2. Core-collapse explosions

The iron stellar cores begin to collapse owing to the loss of dynamical stability. Due to the photo-disintegration of iron into free nucleons and α-particles the adiabatic index γ becomes less than 4/3. Hence, the gradient of pressure cannot withstand the force of gravity any more. An inner core with a mass of $(0.6 - 0.8)$ $M_\odot$ around the stellar center begins to contract almost in a free fall regime. In a few hundredths of second the central density reaches the nuclear density and the contraction slows down. The outermost layers, being still in a state close to free fall, collide with the decelerated inner core. Thereby a nearly steady-state accreting shock wave (SW) forms at the boundary of the inner core and the outer envelope. The key question for the core-collapse supernova mechanism is to verify whether such a steady shock becomes finally transformed into an outgoing blast wave that would expel the supernova envelope. The characteristic features of hydrodynamic flow are schematically shown in Fig. 5.

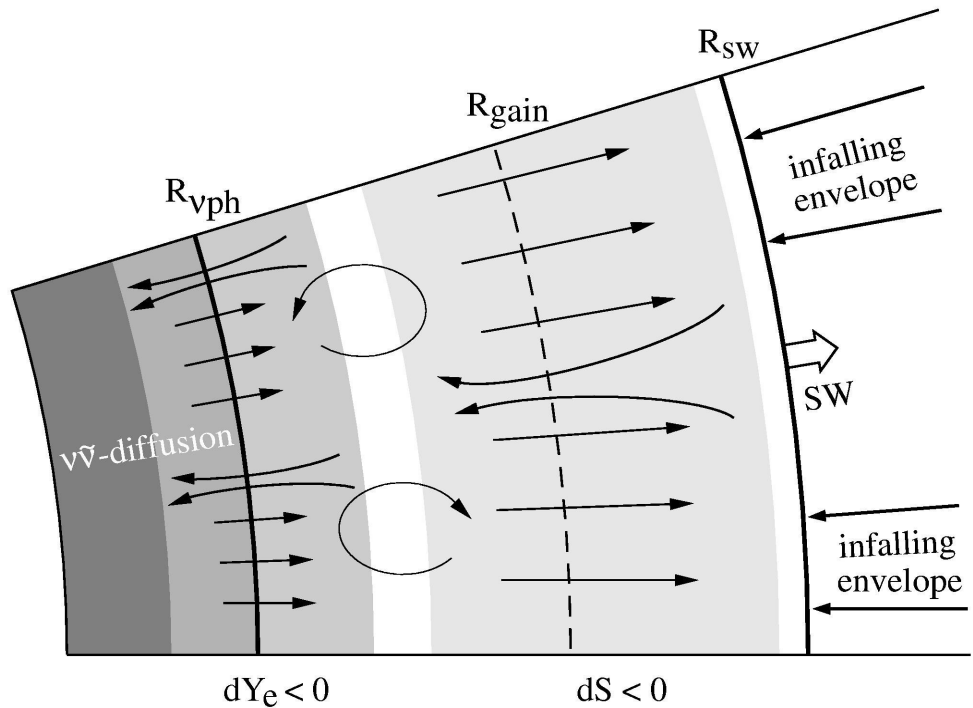

Figure 5. The layout of hydrodynamic flow within the interval between the shock wave (SW) and the neutrinosphere. Two regions of convective instability are shaded in light gray. At radii $r > R_{gain}$ heating of stellar matter by the neutrino flux from the neutrinosphere exceeds its cooling by the local neutrino losses.

The large scale convective currents are expected to evolve from negative gradients of both the entropy (dS < 0, entropy-driven convection) that appears in the region just below the SW and of the electron fraction ($dY_e$ < 0, lepton-driven convection) nearby the neutrinosphere. These currents transport some energy to the SW in addition to that supplied by the neutrino flux from the neutrinosphere. There was suggested that this effect can force the SW to propagate outside.

However, an extensive modeling of the core-collapse SNe during last three decades has demonstrated that neither the convection nor the heating due to the neutrino flux (at $r > R_{gain}$, Fig. 5) can increase the explosion energy to the value that would be large enough to explain the observations. Nevertheless in case of spherically symmetric collapse, the SW finally throws an envelope out. This happens when with time the rate of mass accretion from the envelope considerably decreases and the excessive pressure like an over-compressed elastic spring pushes the SW outward forcing it to propagate through a steep density gradient. Unfortunately this effect (called hydrodynamic bounce) can produce only a low energy explosion with $E_{exp}$ being at least one or even two orders of magnitude less than its standard value of $10^{51}$ erg. Although such a weak explosion seems to be adequate for some subluminous Type II SNe like the historical SN 1054 (see [16] and discussion therein) the problem to get the more energetic explosions remains to be far from a satisfactory solution, at least in the framework of spherically symmetric models. Therefore, it is plausible to assume that the solution can be found by addressing to substantially non-spherical effects such as, for instance, magneto-rotational mechanism [17−19] and rotational fission of the collapsing core into a binary system of proto-neutron stars that evolves losing angular momentum and energy through the gravitational waves with subsequent explosion of a low-mass ( $\approx 0.1 M_{\odot}$ ) component [20, 21].

## 3. Supernova 1987A in the Large Magellanic Cloud

Below we briefly outline the main properties of SN 1987A. Further details and bibliography one can find in a number of comprehensive reviews [22−26]. The most outstanding issues of the SN 1987A are (*i*) the detection of extragalactic neutrinos, (*ii*) the discovery of radioactive nuclides ($^{56}$Ni, $^{56}$Co, $^{57}$Co, $^{44}$Ti) in the SN ejecta, (*iii*) the recognition of the decisive significance of large-scale mixing just before or in the process of the explosion and of aspherical effects, and (*iv*) a colorful picture of the circumstellar environment the progenitor being embedded in.

**Neutrino.** On the day before the optical discovery of SN 1987A February 23, 1987, underground neutrino detectors LSD (Soviet-Italian installation under Mt. Mont Blanc), IMB (Irvine-Michigan-Brookhaven, USA), KII (Kamioka Nuclear Decay Experiment II, Japan), and BUST (Baksan Underground Scintillation Telescope, USSR) had observed a neutrino signal from SN 1987A. The signal consisted of two pulse packets separated by an interval of $4^h 44^m$. The total number of attributed to neutrino events in all the detectors was 8 for the first pulse packet (mainly contributed by LSD) and 28 for the second one (mainly due to IMB and KII). At first glance, the second neutrino pulse nicely confirmed a theoretical prediction of the expected number of events, individual neutrino energies, and pulse duration of (10 − 20) s being determined by cooling of a hot neutron star. However, a close inspection indicated that some properties of the neutrino signal did not allow one to be satisfied with such a coincidence between theory and observations. For instance, for the reaction of the electron antineutrinos with the protons (assumed to be a main source of neutrino signature in the detectors) the momentum of relativistic positrons should not correlate with the direction to SN 1987A, but actually it correlates! The most important result is the presence of two neutrino pulses unexpectedly denoting a two-stage nature of the collapse. An impressive feature of the low-mass neutron star explosion scenario is its ability to explain the two-pulse neutrino signal in the framework of a single self-consistent model.

Theoretical deciphering of the SN 1987A neutrino signal is not yet completed. The crucial point here is a long-awaited discovery of stellar remnant (a neutron star or a black hole) which emergence out of the supernova debris is expected in the near future. At present, only upper limits on the optical and X-ray luminosities of the SN 1987A central region are available [27].

**Radioactive nuclides.** The observations of SN 1987A were the first that gave a decisive experimental evidence of the presence of radioactive nuclides in supernova ejecta. The beta-decay products were identified not only by the slope of the bolometric light curve tail, measured with an unexampled precision, but also by the detection of the X-ray continuum with the *Röntgen* Observatory aboard the *Kvant* module of the *Mir* space station (in the photon energy range 20 – 300 keV) and with Ginga satellite (4 – 30 keV). Moreover, the original (unprocessed as opposed to the X-ray observations) gamma-ray lines from the decay of $^{56}$Co were detected with the *Solar Maximum Mission* satellite and with a number of balloon-borne experiments. The late-time bolometric luminosity of SN 1987A (900 days after the explosion and later on) and the observed hard X-ray imply the energy deposition from 0.075 M$_\odot$ of $^{56}$Ni, a $^{57}$Co/$^{56}$Co between 2.5 and 4 times "solar", and "solar" $^{44}$Ti [28].

**Mixing and asphericity.** A number of features (for instance, such as linear polarization and structure of hydrogen line profiles) seen in detailed high resolution spectra of SN 1987A distinctly denoted violations of strict spherical symmetry of the ejecta (see [29, 30] for the discussion). The closer we look into the center of core-collapse SNe, the stronger are evidences of the asphericity [30]. This suggests that the aspherical effects are perhaps of great importance for the core-collapse supernova mechanism. Early emergence of X-ray from Ni-decay in SN 1987A can be unambiguously explained by a jet-like ejection of a "Ni bullet" of a mass of $\approx 0.0001$M$_\odot$ only [29]. Thus, we see that aspherical effects are directly connected with a large-scale mixing of chemical elements. In particular, in order to achieve an agreement of the light curves obtained by theoretical modeling with those observed for SN 1987A one has to assume that $^{56}$Ni actually was admixed up to large radii where initially it could not be synthesized in thermonuclear reactions [31].

**Circumstellar environment.** The SN 1987A progenitor exploded being in a state of a blue supergiant star. However, some 10,000 years earlier it was passing through the evolutionary stage of a red supergiant losing its mass by a slow $(10 - 30)$ km/s wind. After a short transition phase to the blue supergiant, a fast $\approx 1000$ km/s wind began to blow out circumstellar matter created by the wind of the red one. The interaction between these two winds together with ionizing radiation from the blue supergiant produced a system of three coaxial rings. The progenitor is in the center of the inner ring whereas two other rings are displaced astride along the axis of symmetry. The formation of such a system of rings is described by elegant solutions of equations of hydrodynamics [32, 33]. The rings remained invisible until they were ionized by a flash of ultraviolet radiation from shock wave breakout. In 1999, the fast (moving with a velocity of $\approx 20000$ km/s) outer edge of the SN 1987A reached the inner ring. The resulting collision heated the ring up and it began to shine like a spectacular necklace of glowing knots (see the images provided by the Hubble Space Telescope [34] and further discussion in [35]).

## 4. Conclusion

For lack of space, several important topics were not discussed in this extremely short review. Among them are nucleosynthesis in SNe (in particular, the neutrino-induced nucleosynthesis and r-process in core-collapse SNe), the connection of SNe with gamma-ray bursts and the onset of a neutrino-driven wind that presumably should blow from a nascent hot neutron star. To make up this deficiency we refer the reader to excellent reviews [36–40].

I am very grateful to the Organizing Committee of the Baikal School for warm hospitality and financial support.